\documentclass{article}
\usepackage{times}
\usepackage{amsfonts}
\begin{document}
\title{A Quantum is a Complex Structure\\
on Classical Phase Space}
\author{Jos\'e M. Isidro\\
Instituto de F\'{\i}sica Corpuscular (CSIC--UVEG)\\
Apartado de Correos 22085, Valencia 46071, Spain\\
and\\
Max-Planck-Institut f\"ur Gravitationsphysik, 
\\Albert-Einstein-Institut, \\
D-14476 Golm, Germany\\
{\tt jmisidro@ific.uv.es}}

\maketitle

\begin{abstract}

Duality transformations within the quantum mechanics of a finite number of degrees of freedom can be regarded as the dependence of the notion of a quantum, {\it i.e.}, an elementary excitation of the vacuum, on the observer on classical phase space. Under an observer we understand, as in general relativity, a local coordinate chart. While classical mechanics can be formulated using a symplectic structure on classical phase space, quantum mechanics requires a complex--differentiable structure on that same space. Complex--differentiable structures on a given real manifold are often not unique. This article is devoted to analysing the dependence of the notion of a quantum on the complex--differentiable structure chosen on classical phase space. 


2001 Pacs codes: 03.65.Bz, 03.65.Ca, 03.65.-w.

\end{abstract}

\tableofcontents

\section{Introduction}\label{labastidahijoputa}

\subsection{Summary}\label{cesargomezquetepartaunrayo}

Complex differentiability on a given real manifold often admits several, mathematically nonequivalent definitions. This is of utmost importance for quantum mechanics when the latter is formulated with the aid of classical phase space. Roughly speaking, complex differentiability amounts to a declaration of what depends on $z$ {\it vs.} what depends on $\bar z$, through the Cauchy--Riemann equations. Setting $z=(q+{\rm i}p)/\sqrt{2}$ and $\bar z=(q-{\rm i}p)/\sqrt{2}$ on classical phase space and proceeding to quantisation, $z$ and $\bar z$ respectively become annihilation and creation operators of the quantum theory. Clearly, having more than one possible definition of complex differentiability on classical phase space implies having more than one notion of what an elementary quantum is. This is precisely the concept of a quantum duality \cite{VAFA}.

An important disclaimer is in order. The choice of a complex structure on classical phase space is not new to geometric quantisation \cite{WOODHOUSE}. Thus, {\it e.g.}, in the particular case of Chern--Simons theory \cite{WITTEN}, special effort is devoted to proving the independence of the quantum theory with respect to the choice of a complex structure on classical phase space. In geometric quantisation, independence of the quantum theory with respect to the complex structure can be formulated more or less as follows. There is a bundle of Hilbert spaces of quantum states over a certain base manifold. The latter is the space of all complex structures (satisfying certain natural requirements) that one can place on classical phase space. This bundle of Hilbert spaces admits a projectively flat connection. Being projectively flat, this connection allows for a canonical identification between the fibres corresponding to different choices of a complex structure. 

Geometric quantisation was firmly established already in the 1980's. It never faced the notion of duality, which arose during the second superstring revolution of the mid 1990's \cite{VAFA}. In this paper we interpret dualities as the possibility of having different notions of what an elementary quantum is, depending on the choice of a complex structure on classical phase space. {\it We would like to stress the fact that our conclusions do not clash with geometric quantisation}. Notwithstanding the canonical identification between different fibres alluded to above, our approach to varying the complex structure is entirely different. We perform a canonical quantisation of a number of K\"ahler manifolds (classical phase spaces) whose (local) K\"ahler potentials are taken to be their classical Hamiltonian functions. Solving the corresponding time--independent Schr\"odinger equations, we observe that both the eigenvectors and the eigenvalues exhibit an unambiguous dependence on the complex structure chosen, despite the possibility of parallel--transporting them into those corresponding to another complex structure. As an example, what appears to be a semiclassical state (with respect to a certain complex structure) may well be mapped, by the parallel transport mentioned above, into a highly quantum excitation, as measured by a complex structure that is nonbiholomorphic with the former. Indeed, nonholomorphic maps involving not just $z$ but also $\bar z$, the corresponding quantum creation and annihilation operators are no longer kept separate. One can imagine a transformation mapping, {\it e.g.}, a large number of creation operators (large quantum numbers: the semiclassical regime in terms of the original variables) into a large number of annihilation operators  (small quantum numbers: the strong quantum regime in the new variables). This, of course, is just one conceivable example out of many possible, but it serves well to illustrate our point.

Mathematical background pertaining to the topics analysed here can be found in refs. \cite{KN, GFH, BRUZZO}. References to the quantisation of K\"ahler manifolds, from different standpoints, are \cite{BEREZIN} for the homogeneous case, reviewed in \cite{PERELOMOV} (Berezin quantisation); ref. \cite{SCHLICHENMAIER} for the compact case (Berezin--Toeplitz quantisation), and \cite{TAKHTA} for the general case (deformation quantisation). From the point of view of geometric quantisation, the independence of the quantum theory with respect to the complex structure has been analysed in ref. \cite{WITTEN}; see also \cite{WOODHOUSE}. Although related with our theme, these approaches are entirely different from the viewpoint taken, the techniques applied, and the goals achieved here. We would also like to draw attention to some related papers \cite{COMPEAN, MINIC, MPLA, MATONE, BRODY, ANANDAN, CARROLL, MEX}, where issues partially overlapping with ours are dealt with.

\subsection{Outline}\label{gatitamiau}

In section \ref{labastidamecagoentuputacara} we introduce the Picard group of classical phase space as the parameter space for physically inequivalent vacua. Section \ref{cesargomezchupamelapollaquetefollen} is devoted to analysing the classical and quantum dynamics defined
by the K\"ahler potential, when the K\"ahler potential is chosen to be the classical Hamiltonian. Our analysis is first modelled on the case of the harmonic oscillator on Euclidean space, and then extended (in the appendix) to projective and hyperbolic spaces, as well as some generalisations, namely, Grassmann manifolds and bounded domains. In all these cases the Hilbert space of quantum states is explicitly constructed, by diagonalising the K\"ahler potential, {\it i.e.}, by solving the Schr\"odinger equation. This is done locally over every coordinate patch on which the K\"ahler potential is defined. Whenever the corresponding manifold is not contractible, a set of transition functions are identified in order to define a (possibly nontrivial) bundle of Hilbert spaces over classical phase space. In section \ref{labastidamecagoentuputasombra} we deform the complex structures on the above manifolds and determine the corresponding moduli spaces of complex structures. The same is done in section \ref{mecagoentuputamadrelabastida} for the deformations of the corresponding K\"ahler structures. These deformations are interpreted as dualities of the quantum theory in section \ref{fin}. Explicit examples are presented and worked out in detail in the appendix (section \ref{ketefollenramallodemierda}).

\subsection{Notations}\label{ramalloquetepartaunrayo}

Throughout this article, ${\cal C}$ will denote a real $2n$--dimensional K\"ahler manifold that will play the role of classical phase space.
We will denote the corresponding symplectic form and complex structure by $\omega$ and ${\cal J}$, respectively. Since $\omega$ and ${\cal J}$ are compatible, holomorphic coordinates on ${\cal C}$ will also be Darboux coordinates, up to a possible conformal factor. We will normalise all symplectic forms such that the symplectic volume of ${\cal C}$ equals the dimension of the complex Hilbert space of states obtained by quantisation of ${\cal C}$:
\begin{equation}
\int_{\cal C}\omega^n = {\rm dim}\,{\cal H}.
\label{labastidamariconazo}
\end{equation}
The linear $2n$--dimensional space $\mathbb{R}^{2n}$ has the Darboux coordinates $q^k$, $p^k$, $k=1,\ldots, n$, and the symplectic form
\begin{equation}
\omega=\sum_{k=1}^n{\rm d}q^k\wedge{\rm d}p^k.
\label{hijoputaluisibanez}
\end{equation}
The corresponding quantum operators satisfy 
\begin{equation}
[Q^j,P^k]={\rm i}\delta^{jk}.
\label{kabronluisibanez}
\end{equation}
We endow $\mathbb{R}^n$ with the Euclidean metric 
\begin{equation}
g_{\rm lin}=\frac{1}{2}\sum_{k=1}^n\left(({\rm d}q^k)^2+({\rm d}p^k)^2\right).
\label{putoluisibanez}
\end{equation}
Complex $n$--dimensional space $\mathbb{C}^n$ has the holomorphic coordinates
\begin{equation}
z^k=\frac{1}{\sqrt{2}}\left(q^k+{\rm i}p^k\right), \qquad k=1,\ldots, n,
\label{labastidaquetepartaunrayo}
\end{equation}
and is endowed with the same metric as $\mathbb{R}^{2n}$, now Hermitian instead of real bilinear, 
\begin{equation}
g_{\rm lin}=\sum_{k=1}^n {\rm d}\bar z^k{\rm d}z^k=\vert\vert {\rm d}z\vert\vert^2.
\label{mekagoentuputakaraluisibanez}
\end{equation}
The corresponding quantum operators 
\begin{equation}
A^k=\frac{1}{\sqrt{2}}\left(Q^k+{\rm i}P^k\right),\qquad
(A^k)^+=\frac{1}{\sqrt{2}}\left(Q^k-{\rm i}P^k\right)
\label{luisibanezquetepartaunrayo}
\end{equation}
satisfy
\begin{equation}
[A^j,(A^k)^+]=\delta^{jk}.
\label{luisibanezmekagoentuputakaramarikondemierda}
\end{equation}
On a general K\"ahler phase space ${\cal C}$ other than $\mathbb{C}^n$, eqns. (\ref{hijoputaluisibanez}) to (\ref{luisibanezmekagoentuputakaramarikondemierda}) will hold locally on every coordinate chart, possibly up to conformal factors.

\section{Varying the vacuum}\label{labastidamecagoentuputacara}

Given a complex manifold ${\cal C}$, complex line bundles over it can be arranged into holomorphic equivalence classes, and the quotient set can be given a group structure. The result is the Picard group of ${\cal C}$, denoted ${\rm Pic}\,({\cal C})$ \cite{GFH}. The way in which ${\rm Pic}\,({\cal C})$ enters the quantum theory is as follows.

Assume that ${\cal C}$ is covered by a collection of holomorphic coordinate charts $({\cal U}_j, z^k_{(j)})$, where $k=1, \ldots, n$ runs over the complex dimensions of ${\cal C}$ and $j$ runs over all the charts in a holomorphic atlas. As in ref. \cite{PQM} we erect, on every chart $({\cal U}_j, z^k_{(j)})$, a vector--space fibre that will play the role of the Hilbert space of states; this is done by identifying a vacuum state $\vert 0(j)\rangle$ and acting on it with (products of powers of) the creation operators $(A^k(j))^+$. Assuming initially that the vacuum is nondegenerate, every choice of a set of holomorphic transition functions for the complex vector $\vert 0(j)\rangle$ defines a complex line bundle on ${\cal C}$ as we cover the latter with coordinate charts $({\cal U}_j, z^k_{(j)})$. Hence the vacuum state defines a class of holomorphic vector bundles over ${\cal C}$, {\it i.e.}, an element of ${\rm Pic}\,({\cal C})$. Conversely, every class in ${\rm Pic}\,({\cal C})$ determines a holomorphic line bundle, whose fibrewise generator (a complex vector) we can take to be the vacuum state on each coordinate chart. The action of creation operators on this vector gives rise to excitations of the vacuum, {\it i.e.}, quantum states.

Degenerate vacua can be treated similarly. Assume that we have a $d$--fold degenerate vacuum, spanned on the chart $({\cal U}_j, z^k_{(j)})$ by the vectors $\vert 0(j)_1\rangle$, $\ldots$, $\vert 0(j)_d\rangle$. By the assumption of degeneracy, they are all physically indistinguishable, so it makes sense to consider their wedge product $\wedge_{m=1}^d\vert 0(j)_m\rangle$, which changes at most by a sign under permutations of its $d$ factors.  Taking this wedge product to be the fibrewise generator of a line bundle over ${\cal C}$, we determine a class in ${\rm Pic}\,({\cal C})$. Now not every class in ${\rm Pic}\,({\cal C})$ gives rise to a $d$--fold degenerate vacuum, as the transition functions must be such that a $d$--th root must exist, so that this root also defines a set of transition functions for each and every one of the $d$ vectors $\vert 0(j)_m\rangle$. The assumption of degeneracy implies that the same set of transition functions must be valid for all $m=1, \ldots, d$. This can be ensured by taking the parameter space for inequivalent $d$--fold degenerate vacua to be the $d$--th power of the Picard group ${\rm Pic}\,({\cal C})$. That is, we take the $d$--th power of all classes in ${\rm Pic}\,({\cal C})$, as their $d$--th root is then (and only then) a true class. The resulting set provides the correct parameter space for degenerate vacua.

\section{K\"ahler manifolds as classical phase spaces}\label{cesargomezchupamelapollaquetefollen}

The simplest K\"ahler manifold is the linear space $\mathbb{C}^n$. A possible K\"ahler potential is
\begin{equation}
K_{\rm lin}=\vert\vert z\vert\vert^2,
\label{putonbarbon}
\end{equation}
the K\"ahler symplectic form being
\begin{equation}
\omega_{\rm lin}=-{\rm i}\sum_{k=1}^n{\rm d}{\bar z}^k\wedge{\rm d}z^k.
\label{barbonputon}
\end{equation}
The symplectic volume of $\mathbb{C}^n$ is infinite,
\begin{equation}
\int_{\mathbb{C}^n}\omega_{\rm lin}^n=\infty.
\label{kagonlag}
\end{equation}
$\mathbb{C}^n$ being contractible, it can be covered with a single coordinate chart (the $z^k$ above),
so all vector bundles over $\mathbb{C}^n$ are necessarily trivial. In particular its Picard group is trivial,
\begin{equation}
{\rm Pic}\,(\mathbb{C}^n)=0.
\label{cesargomezchupamelapolla}
\end{equation}
We denote by $\vert 0\rangle_{\rm lin}$ the fibrewise generator of the trivial complex line bundle over $\mathbb{C}^n$.
Now the classical Hamiltonian function $H_{\rm lin}$ on $\mathbb{C}^n$ equals the K\"ahler potential (\ref{putonbarbon}), $H_{\rm lin}=K_{\rm lin}$. 
This is the dynamics of the $n$--dimensional linear harmonic oscillator, whose canonical equations of motion read
\begin{equation}
\dot z^k=-{\rm i}\,\frac{\partial K_{\rm lin}}{\partial \bar z^k}=-{\rm i}\, z^k.
\label{delaguilaeresunpedazodeimbecilquetecagas}
\end{equation}
The quantisation of this dynamics is well known. The classical coordinates $z^k$ and their complex conjugates $\bar z^k$ respectively give rise to annihilation and creation operators $A^k_{\rm lin}$ and $(A^k_{\rm lin})^+$ acting on the vacuum $\vert 0\rangle_{\rm lin}$. The quantum Hamiltonian operator is 
\begin{equation}
H_{\rm lin}=\sum_{k=1}^n\left((A^k_{\rm lin})^{+} A^k_{\rm lin}+{1\over 2}\right),
\label{cesargomezhijoputa}
\end{equation}
and its eigenvalue equation
\begin{equation}
\sum_{k=1}^n\left((A^k_{\rm lin})^{+} A^k_{\rm lin}+{1\over 2}\right)|m_1,\ldots, m_n\rangle_{\rm lin} =
E_n|m_1,\ldots, m_n\rangle_{\rm lin}
\label{velascocabronazo}
\end{equation}
is solved by the eigenvalues $E_n=\sum_{k=1}^n \left(m_k + {1\over 2}\right)$, with the eigenstates
\begin{equation}
|m_1,\ldots, m_n\rangle_{\rm lin} = \frac{1}{\sqrt{m_1!\cdots m_n!}}
\left((A^1_{\rm lin})^+\right)^{m_1}\cdots \left((A^n_{\rm lin})^+\right)^{m_n}\vert 0\rangle_{\rm lin},
\label{velascocabron}
\end{equation}
which are excitations of the vacuum $\vert 0\rangle_{\rm lin}$. The Hilbert space ${\cal H}_{\rm lin}$ is (the closure of) the linear span of all the states $|m_1,\ldots, m_n\rangle_{\rm lin}$, where the occupation numbers $m_k$ $k=1,\ldots, n$, run over all the nonnegative integers.
Thus ${\cal H}_{\rm lin}$ is infinite--dimensional, in agreement with eqns. (\ref{labastidamariconazo}), (\ref{kagonlag}).

The previous results can be extended to more general K\"ahler manifolds. As before, let us consider a K\"ahler manifold ${\cal C}$ covered by a holomorphic atlas with coordinate charts $({\cal U}_j, z^k_{(j)})$. For simplicity we will drop the subindex $j$ from our notations, bearing in mind, however, that we are working locally on the $j$--th chart. On the latter, K\"ahler potentials $K(\bar z^k, z^k)$ are defined only up to gauge transformations 
\begin{equation}
K(\bar z^k, z^k)\longrightarrow K(\bar z^k, z^k)+F(z^k)+G(\bar z^k),
\label{alvarezgaumeeresuncasposodemierda}
\end{equation}
where $F(z^k)$ is an arbitrary holomorphic function and $G(\bar z^k)$ an arbitrary antiholomorphic function on the given chart. Hence terms depending exclusively on $z^k$ or exclusively on $\bar z^k$ can be gauged away. With this choice of gauge the K\"ahler potential on the given chart is unique, given that its overall normalisation is fixed by eqn. (\ref{labastidamariconazo}). Such a potential always exists locally on a K\"ahler manifold ${\cal C}$; it is a real, smooth function that factorises as the product of a holomorphic function $F(z^k)$ times an antiholomorphic function $G(\bar z^k)$,
\begin{equation}
K(z^k, \bar z^k)=F(z^k)G(\bar z^k),
\label{luisibanezmecagoentuputakaramarikondemierda}
\end{equation}
or, more generally, as a sum of such terms. In general, however, no K\"ahler potential can be defined {\it globally}\/ on ${\cal C}$, and cases like $\mathbb{C}^n$, where a global K\"ahler potential {\it does}\/ exist, are rather exceptional. A nontrival de Rham cohomology group $H^2({\cal C}, \mathbb{R})$ is an obstruction to the existence of a globally--defined K\"ahler potential \cite{KN, GFH}. 

Now let us assume that, on every coordinate chart $({\cal U}_j, z^k_{(j)})$, a K\"ahler potential can be found that is a function of $\vert\vert z\vert\vert^2=\sum_{k=1}^n\bar z^kz^k$,
\begin{equation}
K(\bar z^k, z^k)=K_{\cal C}(\vert\vert z\vert\vert^2).
\label{cesargomezeresuncasposo}
\end{equation}
Now for the potential (\ref{luisibanezmecagoentuputakaramarikondemierda}) to satisfy the requirement (\ref{cesargomezeresuncasposo}) it is necessary and sufficient that it be $U(n)$--invariant. Indeed, given $U\in U(n)$, consider the coordinates $z^k$ as a column vector and their complex conjugates $\bar z^k$ as a row vector, with the $z^k$ transforming into $U^k_mz^m$ and the $\bar z^k$ into $\bar z^m U_m^k$. Then $\vert\vert z\vert\vert^2$ (or functions thereof) is the unique $U(n)$--invariant one can build. This assumption rules out potentials like, {\it e.g.}, $z^2\bar z + \bar z^2 z$, whose summands are {\it unbalanced}, so to speak, in their holomorphic/antiholomorphic dependence. This gives a K\"ahler metric
\begin{equation}
g_{km}=\frac{\partial^2K_{\cal C}(\vert\vert z\vert\vert^2)}{\partial\bar z^k\partial z^m}.
\label{kagontodos}
\end{equation}
In sections \ref{alvarezgaumemekagoentuputamadremarikonazodeplaya} and \ref{delaguilaeresuncasposodemierda} we will generalise this dependence of the K\"ahler potential on $\vert\vert z\vert\vert^2$ to a dependence on ${\rm tr}(z^+z)$, where $z$ will be a matrix--valued coordinate. As a second assumption we will suppose that $K_{\cal C}(\vert\vert z\vert\vert^2)$, and its generalisation $K_{\cal C}({\rm tr}(z^+z))$ of sections
\ref{alvarezgaumemekagoentuputamadremarikonazodeplaya} and \ref{delaguilaeresuncasposodemierda}, are real--analytic functions in their respective arguments.

Under the above assumptions, we define a dynamics on ${\cal C}$ by taking the classical Hamiltonian function on the coordinate chart $({\cal U}_j, z^k_{(j)})$ equal to the K\"ahler potential on the same chart,
\begin{equation}
H_{\cal C}=K_{\cal C}.
\label{delaguilakabron}
\end{equation}
Now eqn. (\ref{delaguilakabron}) defines only a {\it local}\/ dynamics on ${\cal C}$ since, as explained above, a general ${\cal C}$ admits no globally--defined potential. More importantly, implicitly contained in eqn. (\ref{delaguilakabron}) is the following statement: the space of all solutions to Hamilton's classical equations of motion with respect to the Hamiltonian (\ref{delaguilakabron}) is the manifold ${\cal C}$ itself. 
Thus our choice (\ref{delaguilakabron}) makes sense, because classical phase space is in fact the space of all solutions to Hamilton's equations (modulo possible gauge symmetries). Finally, the extrema of $H_{\cal C}$ will always be minima, as follows from the positivity of the metric (\ref{kagontodos}). Thus picking a Hamiltonian equal to the K\"ahler potential is physically sound.

In order to quantise the dynamics (\ref{delaguilakabron}) on the coordinate chart $({\cal U}_j, z^k_{(j)})$ we will conformally transform the K\"ahler metric (\ref{kagontodos}) into the Euclidean metric $\sum_{k=1}^n{\rm d}\bar w^k{\rm d}w^k$ by means of a coordinate transformation
\begin{equation}
z^k\rightarrow w^k(\bar z^m, z^m).
\label{queputocalvoeresibanez}
\end{equation}
Next we will replace the classical function $\vert\vert w\vert\vert^2$ with the operator of eqn. (\ref{cesargomezhijoputa}),
\begin{equation}
\vert\vert w\vert\vert^2\mapsto H_{\rm lin}=\sum_{k=1}^n\left((A^k_{\rm lin})^+A^k_{\rm lin}+\frac{1}{2}\right).
\label{casposostodos}
\end{equation}
This quantisation prescription also carries a choice of operator ordering attached to it. Then the K\"ahler potential gives rise to a quantum Hamiltonian operator whose diagonalisation, in principle, can be performed using eqns. (\ref{cesargomezhijoputa})--(\ref{velascocabron}). In this way we can erect, over each coordinate chart $({\cal U}_j, z^k_{(j)})$ on ${\cal C}$, a vector--space fibre given by the Hilbert space of quantum states. However, in order to obtain the complete quantum theory, one also needs the following elements: {\it i)} the precise conformal transformation (\ref{queputocalvoeresibanez}); {\it ii)} a set of transition functions for patching together the Hilbert--space fibres across overlapping charts (when ${\cal C}$ is not contractible); {\it iii)} the Picard group ${\rm Pic}\,({\cal C})$ (when ${\cal C}$ is not contractible); {\it iv)} the symplectic volume of ${\cal C}$. These points are best illustrated with the examples given in the appendix (section \ref{ketefollenramallodemierda}).

\section{Complex--structure deformations}\label{labastidamecagoentuputasombra}

The analysis of sections \ref{labastidamecagoentuputacara} and \ref{cesargomezchupamelapollaquetefollen} assumes that a complex structure has been picked on ${\cal C}$ and kept fixed throughout. However one can also consider varying the complex structure on classical phase space. 

Let us first consider $\mathbb{C}^n$. It has a moduli space of complex structures that are compatible with a given orientation \cite{DOUADY}. This moduli space is denoted ${\cal M}(\mathbb{C}^n)$; it is the symmetric space
\begin{equation}
{\cal M}(\mathbb{C}^n)=SO(2n)/U(n).
\label{amico}
\end{equation}
This is a compact space of real dimension $n(n-1)$. Here the embedding of $U(n)$ into $SO(2n)$ is given by
\begin{equation}
A+{\rm i}B\longrightarrow\left(\begin{array}{cc}
A&B\\
-B&A
\end{array}\right),
\label{labastidaquetelametanporculo}
\end{equation}
where $A+{\rm i}B\in U(n)$ with $A, B$ real, $n\times n$ matrices \cite{HELGASON}. 

Let us see how the symmetric space (\ref{amico}) appears as a moduli space \cite{DOUADY, HELGASON}. Consider the Euclidean metric $g_{\rm lin}$ of eqn. (\ref{putoluisibanez}). Requiring rotations to preserve the orientation, the isometry group of $g_{\rm lin}$ is $SO(2n)$. In the complex coordinates of eqn. (\ref{labastidaquetepartaunrayo}), $g_{\rm lin}$ becomes the Hermitian form (\ref{mekagoentuputakaraluisibanez}), whose isometry group is $U(n)$. Notice that we no longer impose the condition of unit determinant, since 
$U(n)=SU(n)\times U(1)$ and $g_{\rm lin}$ is invariant under the $U(1)$ action $z^k\rightarrow {\rm e}^{{\rm i}\alpha}z^k$, $k=1,\ldots, n$, for all $\alpha\in\mathbb{R}$. Now every choice of orthogonal axes $x^k$, $y^k$ in $\mathbb{R}^{2n}$, {\it i.e.}, every element of $SO(2n)$, defines a complex structure on $\mathbb{R}^{2n}$ upon setting 
\begin{equation}
w^k=\frac{1}{\sqrt{2}}\left(x^k+{\rm i}y^k\right),\qquad k=1,\ldots, n. 
\label{kakaxramallo}
\end{equation}
Generically the $w^k$ are related nonbiholomorphically with the $z^k$, because the orthogonal transformation 
\begin{eqnarray}
z^k\longrightarrow w^k&=&R^k_mz^m+S^k_{ m}\bar z^{ m}\nonumber\\
\bar z^{ k}\longrightarrow \bar w^{ k}&=&\bar R^{ k}_{ m}\bar z^{ m}+\bar S^{k}_{m} z^{m},
\label{mierdaxramallo}
\end{eqnarray}
while satisfying the orthogonality conditions
\begin{equation}
R^k_m\,\bar R^{ k}_{ n}+S^k_{ n}\, \bar S^{ k}_m=\delta_{mn},\qquad
R^k_m\,\bar S^{ k}_n=0=S^k_{ m}\,\bar R^{ k}_{ n},
\label{kakaxbarbon}
\end{equation}
need not satisfy the Cauchy--Riemann conditions 
\begin{equation}
\frac{\partial \bar w^{ k}}{\partial z^m}=\bar S^{ k}_m=0=S^k_{ m}=\frac{\partial  w^{k}}{\partial \bar z^{m}}.
\label{mierdaxcesargomez}
\end{equation}
However, when eqn. (\ref{mierdaxcesargomez}) holds, the transformation (\ref{mierdaxramallo}) is not just orthogonal but also unitary. 
Therefore one must divide $SO(2n)$ by the action of the unitary group $U(n)$, in order to obtain the parameter space for rotations that truly correspond to inequivalent complex structures on $\mathbb{R}^{2n}\simeq \mathbb{C}^n$. Nonbiholomorphic complex structures on $\mathbb{C}^n$ are  1--to--1 with rotations of $\mathbb{R}^{2n}$ that are {\it not}\/ unitary transformations.

When $n=1$ the moduli space (\ref{amico}) reduces to a point. Therefore on the complex plane $\mathbb{C}$ there exists a unique complex structure, that we can identify as the one whose holomorphic atlas consists of the open set $\mathbb{C}$ endowed with the holomorphic coordinate $z=(q+{\rm i}p)/\sqrt{2}$. Physically this corresponds to the 1--dimensional harmonic oscillator. Consider now $n$ independent harmonic oscillators, where ${\cal C}=\mathbb{C}^n=\mathbb{C}\times{n\atop\cdots}\times\mathbb{C}$. Although it is never explicitly stated, the complex structure on this product space is always understood to be the $n$--fold Cartesian product of the unique complex structure on $\mathbb{C}$. Obviously, removing the requirement of compatibility between the complex structure and the orientation chosen, we duplicate the number of complex structures.

Let us finally analyse the complex--structure moduli of projective and hyperbolic spaces (see appendix). For projective space we have $\mathbb{CP}^n=\mathbb{C}^n\cup \mathbb{CP}^{n-1}$, with $\mathbb{CP}^{n-1}$ a hyperplane at infinity. It follows that ${\cal M}(\mathbb{CP}^n)={\cal M}(\mathbb{C}^n)$. The case $n=1$ is interesting because $\mathbb{CP}^1=S^2$, the Riemann sphere. The latter can be regarded as the classical phase space of a spin--1/2 system, inasmuch as spin possesses a classical counterpart. Hence there are no complex--structure moduli on the Riemann sphere. For hyperbolic space we also have ${\cal M}(B^n)={\cal M}(\mathbb{C}^n)$, since the complex structure on $B^n$ is the one induced by $\mathbb{C}^n$. Grassmann manifolds $G_{r,r'}(\mathbb{C})$ and bounded domains $D_{r,r'}(\mathbb{C})$ are natural generalisations of projective and hyperbolic space, respectively, so analogous conclusions apply to them.

\section{K\"ahler deformations}\label{mecagoentuputamadrelabastida}

Next we study the dependence of the quantum theory on the K\"ahler moduli, while keeping the complex moduli fixed, in the cases when ${\cal C}$ is linear space $\mathbb{C}^n$, hyperbolic space $B^n$ and projective space $\mathbb{CP}^n$. We will show that these 3 cases correspond to  different approximation regimes of the K\"ahler class.

Let us first consider the restriction of the K\"ahler potential (\ref{putonbarbon}) to the unit ball $B^n$, and let us deform it by a polynomial of degree $N>1$, 
\begin{equation}
K_{\rm lin}\rightarrow K_{(N)}=\vert\vert z\vert\vert^2+\frac{1}{2}\vert\vert z\vert\vert^4+\frac{1}{3}\vert\vert z\vert\vert^6+\ldots+\frac{1}{N}\vert\vert z\vert\vert^{2N}.
\label{barbonmaricon}
\end{equation}
This deformation gives rise to a new K\"ahler potential on $B^n$. Let $\omega_{(N)}$ denote the deformed symplectic form corresponding to $K_{(N)}$, and let $B_{(N)}^n$ denote the resulting manifold, with $B^n_{(N=1)}=B^n$.  Any deformation of finite degree $N$ increases the symplectic volume by a finite amount. This increase is positive  because all summands in eqn. (\ref{barbonmaricon}) are positive definite. Despite its increase, the symplectic volume of $B^n_{(N)}$ measured by the $2n$--form $\omega_{(N)}^{n}$ always remains finite:
\begin{equation}
\int_{B^n_{(N)}}\omega_{(N)}^n<\infty,\qquad 1<N<\infty.
\label{cesargomezcabron}
\end{equation}
{}For all finite $N>1$, $K_{(N)}$ is a K\"ahler deformation of $K_{\rm lin}$ that increases the symplectic volume of $B^n$. Then eqns. (\ref{velascocabronazo}), (\ref{velascocabron}) allow one to diagonalise the Hamiltonian
\begin{equation}
H_{(N)}=\sum_{j=1}^N\frac{1}{j}\left(\sum_{k=1}^n\left((A^k_{\rm lin})^+A^k_{\rm lin}+\frac{1}{2}\right)\right)^{j}.
\label{velascomaricon}
\end{equation}
It has eigenstates $\vert m_1,\ldots,m_n\rangle_{\rm lin}$ corresponding to the (nondegenerate) eigenvalues
\begin{equation}
\sum_{j=1}^N\frac{1}{j}\left(\sum_{k=1}^n\left(m_k+\frac{1}{2}\right)\right)^{j},
\label{velascomaricas}
\end{equation}
where the occupation numbers $m_k$ do not run over all the nonnegative integers: they are limited by the constraint (\ref{cesargomezcabron}) to a finite range. Although the precise value of this range is immaterial for our purposes, let us say that it can be determined (up to irrelevant multiplicative factors) as the whole part of the integral (\ref{cesargomezcabron}); as such it is a positive, monotonically increasing function of $N$, divergent in the limit $N\rightarrow \infty$ where, thanks to the Taylor expansion ($x=\vert\vert z\vert\vert^2$)
\begin{equation}
-\ln(1-x)=x+\frac{1}{2}x^2+\frac{1}{3}x^3+\frac{1}{4}x^4+\ldots \qquad \vert x\vert<1,
\label{cagonramallo}
\end{equation}
the results of section \ref{javiermaschupamelapolla} are reproduced. In the limit $N\to\infty$ the manifold $B^n_{(N)}$ becomes the hyperbolic manifold $B_{\rm hyp}^n$, and eqns. (\ref{velascomaricon}), (\ref{velascomaricas}) become their hyperbolic partners (\ref{hacheproj}), (\ref{oeggppv}); the function $h_{\rm hyp}$ of eqn. (\ref{putodelaguila}) appears in the process.
Thus the effect of the K\"ahler deformation (\ref{barbonmaricon}) is that of enlarging the Hilbert space, allowing for excitations of the vacuum obtained by the action of more than just one creation operator $(A^k_{\rm lin})^{+}$. Analogous conclusions would hold if we considered arbitrary positive coefficients $c_j$ multiplying the deformations $\vert\vert z\vert\vert^{2j}$ in eqn. (\ref{barbonmaricon}), instead of $c_j=1/j$. 

Choices for the $c_j$ not all positive, such as $c_j=(-1)^{j+1}/j$, lead to different deformations of the K\"ahler potential (\ref{putonbarbon}) on $\mathbb{C}^n$:
\begin{equation}
K_{\rm lin}\rightarrow K_{(N)}=\vert\vert z\vert\vert^2-\frac{1}{2}\vert\vert z\vert\vert^4+\frac{1}{3}\vert\vert z\vert\vert^6-\ldots+\frac{(-1)^{N+1}}{N}\vert\vert z\vert\vert^{2N}.
\label{barbonmarikon}
\end{equation}
In the limit $N\to\infty$, apply the Taylor expansion ($x=\vert\vert z\vert\vert^2$)
\begin{equation}
\ln(1+x)=x-\frac{1}{2}x^2+\frac{1}{3}x^3-\frac{1}{4}x^4+\ldots \qquad \vert x\vert<1,
\label{casposoramallo}
\end{equation}
initially only to the manifold $B^n$ for convergence. Once the series (\ref{casposoramallo}) has been summed up, take the left--hand side as the K\"ahler potential on all of $\mathbb{C}^n$, and declare the latter to be just one of the $n+1$ coordinate charts on $\mathbb{CP}^n$ described in section \ref{labastidakabron}. Then this deformation of the linear dynamics reproduces the projective dynamics of section \ref{labastidakabron}.

Conversely, taking due care of the domains for the coordinate charts, the hyperbolic and projective dynamics of sections \ref{javiermaschupamelapolla} and \ref{labastidakabron} can be linearised, as per eqns. (\ref{cagonramallo}), (\ref{casposoramallo}), respectively, in order to yield the linear dynamics of section \ref{cesargomezchupamelapollaquetefollen}. In this way the effect of varying the K\"ahler moduli is to deform the symplectic volume of ${\cal C}$. By eqn. (\ref{labastidamariconazo}), this is reflected as a variation in the number of quantum states.

The moduli space of K\"ahler structures on $\mathbb{CP}^n$ is $\mathbb{R}^+$, {\it i.e.}, the positive reals. All these  K\"ahler deformations are compatible with the fixed complex structure. (K\"ahler moduli are associated with what in ref. \cite{PQM} we called {\it representations}\/ for $\mathbb{CP}^n$).

\section{Discussion}\label{fin}

Complex--differentiable structures on classical phase spaces ${\cal C}$ have a twofold meaning. Geometrically they define  
complex differentiability, or analyticity, of functions on complex manifolds such as ${\cal C}$. Quantum--mechanically they  
define the notion of a quantum, {\it i.e.}, an elementary excitation of the vacuum state. In this paper we have elaborated on this latter meaning. The mathematical possibility of having two or more nonbiholomorphic complex--differentiable structures on a given classical phase space leads to the physical notion of a quantum--mechanical duality, {\it i.e.}, to the relativity of the notion of an elementary quantum. This relativity is understood as the dependence of the notion of a quantum on the choice of a complex--differentiable structure on ${\cal C}$. We have summarised this fact in the statement that a quantum is a complex--differentiable structure on classical phase space.

In this article we have proposed a solution to the problem suggested in ref. \cite{VAFA}, namely, how to implement duality transformations in the quantum mechanics of a finite number of degrees of freedom. We have first drawn attention to the key role that complex--differentiable structures on classical phase space play in the formulation of quantum mechanics, without resorting to geometric quantisation. This raises the question, how does quantum mechanics vary with each choice of a complex structure on classical phase space? What does it mean, to have different possible quantum--mechanical descriptions of a given physics? We claim that there are at least three ways in which one can obtain different quantum--mechanical theories over a given classical phase space, thus giving rise to dualities:\\ 
{\it i)} by varying the ground state, {\it i.e.}, the vacuum;\\ 
{\it ii)} by varying the type of excitations of the vacuum, {\it i.e.}, the creation and annihilation operators;\\ 
{\it iii)} by varying the number of excitations of the vacuum, {\it i.e.}, the dimension of the Hilbert space of quantum states.

Each one of these variations, while referring to the quantum theory, concerns properties of classical phase space.
Moreover, each one of these variations has its own parameter space.  The parameter space for physically inequivalent vacua is the Picard group of classical phase space (section \ref{labastidamecagoentuputacara}). The parameter space for physically inequivalent pairs of creation and annihilation operators is the moduli space of complex structures on classical phase space (section \ref{labastidamecagoentuputasombra}).  The parameter space for the dimension of the Hilbert space of states is the moduli space of K\"ahler classes on classical phase space (section \ref{mecagoentuputamadrelabastida}). 

On $\mathbb{C}^n$ every complex structure induces a compatible symplectic structure, by taking the real and imaginary parts of 
$z^k=(q^k+{\rm i}p^k)/\sqrt{2}$ as Darboux coordinates.  On $\mathbb{C}^n$ the converse is also true: Darboux coordinates can be arranged into the real and imaginary parts of holomorphic coordinates. Hence, on $\mathbb{C}^n$, there is a 1--to--1 correspondence between complex structures and  symplectic structures, and a variation in one of them induces a corresponding variation in the other. Differences in the notion of an elementary quantum on $\mathbb{C}^n$ can therefore be traced back to different choices of the classical symplectic structure. Moreover, the Picard group of $\mathbb{C}^n$ being trivial, the corresponding vacuum is also unique (for each choice of a complex structure). Altogether there is no room for dualities on $\mathbb{C}^n$. 

However, on other classical phase spaces (see appendix) there is room for independent variations of complex and symplectic structures, and/or for choosing physically nonequivalent vacua. We have shown that, on the unit ball $B^n\subset \mathbb{C}^n$, we can deform the symplectic structure while keeping the complex structure fixed. This is not quite a quantum--mechanical duality yet, as the quantum theory refers to a complex structure, but further examples can be manufactured. Thus, on the complex 1--dimensional torus one can vary the complex structure while keeping the symplectic structure fixed \cite{TORUS}. This is an example of a quantum--mechanical duality that passes completely unnoticed at the classical level, as it leaves the symplectic structure unchanged. On complex projective space $\mathbb{CP}^n$ there is a nontrivial Picard group, which allows for different vacua \cite{PQM}. 

A duality arises as the possibility of having two or more, apparently different, quantum--mechanical descriptions of the same physics. Above we have enumerated three possible ways in which one can vary the description of a given physics. These facts imply that the concept of a quantum is not absolute, but relative to the quantum theory used to measure it \cite{VAFA}. That is, duality expresses the relativity of the concept of a quantum. In particular {\it classical}\/ and {\it quantum}\/, for long known to be intimately related, are not necessarily always the same for all observers on phase space. 

When ${\cal C}$ is not only complex but also K\"ahler, we have a natural arena for the study of quantum--mechanical dualities. A (local) classical Hamiltonian function can always be found, namely the K\"ahler potential, such that the corresponding canonical equations of motion have ${\cal C}$ as the space of all solutions. We have quantised this classical dynamics by means of a change of variables that essentially reduces the problem to a variant of the harmonic oscillator on Euclidean space $\mathbb{C}^n$ (itself the simplest K\"ahler manifold). Now K\"ahler spaces typically have complex--structure deformation moduli as well as K\"ahler--deformation moduli. We have argued that moving around in their respective moduli spaces, {\it i.e.}, varying these moduli, we obtain different quantum--mechanical descriptions of a given physics. This is precisely the notion of a  quantum duality \cite{VAFA}. 

It is important to stress that our interpretation of the dependence of the quantum theory on the complex structure on classical phase space does not clash with established knowledge. Geometric quantisation \cite{WOODHOUSE, WITTEN} precedes the notion of duality \cite{VAFA} by many years. At the time when geometric quantisation was being developed there was no need to consider variations in the complex structure.  Indeed the mere possibility of having to pick one particular complex structure was somewhat embarrrasing, as one then had to justify the choice of an unwelcome mathematical entity, or at least prove the independence of the quantum theory on the choice made. In the light of developments in string duality and M--theory, the notion of duality gives the choice of a complex structure on classical phase space an interesting physical interpretation.

\section{Appendix: examples of dynamics on K\"ahler spaces}\label{ketefollenramallodemierda}

A number of examples are worked out explicitly in this section.

\subsection{Hyperbolic space}\label{javiermaschupamelapolla}

Within $\mathbb{C}^n$ we have the unit ball
\begin{equation}
B^n=\{(z^1,\ldots, z^n)\in \mathbb{C}^n: \Vert z\vert\vert < 1\}.
\label{cagonbarbon}
\end{equation}
Consider the K\"ahler potential on $B^n$
\begin{equation}
K_{\rm hyp}=-\ln\left(1-\vert\vert z\vert\vert^2\right),
\label{cagonlabastida}
\end{equation}
from which the hyperbolic symplectic form 
\begin{equation}
\omega_{\rm hyp} = {-{\rm i}\over (1-\vert\vert z\vert\vert^2)^2}\sum_{k=1}^n{\rm d} {\bar  z}^k\wedge{\rm d} z^k
\label{sonrisitaslabastida}
\end{equation}
and the hyperbolic metric
\begin{equation}
g_{\rm hyp} = {1\over (1-\vert\vert z\vert\vert^2)^2}\sum_{k=1}^n{\rm d} {\bar  z}^k{\rm d} z^k
\label{sonrisitaslabastidaquetehostienmarikon}
\end{equation}
follow. Hyperbolic space is the K\"ahler manifold obtained by endowing the unit ball (\ref{cagonbarbon}) with the 
K\"ahler potential (\ref{cagonlabastida}). It has constant negative scalar curvature. The symplectic volume of $B^n$ is infinite,
\begin{equation}
\int_{B^n}\omega^n_{\rm hyp}=\infty.
\label{ramallomecagoentuputamadre}
\end{equation}
$B^n$ is contractible. Hence it can be covered with a single coordinate chart (the $z^k$ above), and all vector bundles over $B^n$ are trivial. In particular its Picard group is trivial,
\begin{equation}
{\rm Pic}\,(B^n)=0.
\label{ramayoquetepartaunrayo}
\end{equation}
Let $\vert 0\rangle_{\rm hyp}$ denote the (fibrewise) generator of the trivial complex line bundle over $B^n$. We take the classical Hamiltonian function on $B^n$ equal to the K\"ahler potential (\ref{cagonlabastida}). Let $\pi^{rs}_{\rm hyp}$ denote the Poisson tensor corresponding to the symplectic form $\omega_{\rm hyp}$ of eqn. (\ref{sonrisitaslabastida}). Now one can verify that the space of all solutions to Hamilton's equations
\begin{equation}
\dot z^k=\left\{z^k,K_{\rm hyp}\right\}=\pi^{rs}_{\rm hyp}\frac{\partial z^k}{\partial z^r}\frac{\partial K_{\rm hyp}}{\partial \bar z^s}=
\pi^{ks}_{\rm hyp}\frac{\partial K_{\rm hyp}}{\partial \bar z^s}=-{\rm i}\,z^k\left(1-\vert\vert z\vert\vert^2\right)
\label{velascoquetepartaunrayo}
\end{equation}
is in fact $B^n$. On the latter manifold the Hamiltonian (\ref{cagonlabastida}) is bounded from below, as expected physically.
The right--hand side of the equations of motion (\ref{velascoquetepartaunrayo}) contains the square root of the conformal factor 
\begin{equation}
f_{\rm hyp}=\left(1-\vert\vert z\vert\vert^2\right)^2 
\label{labastidaquetehostien}
\end{equation}
needed to transform the hyperbolic metric (\ref{sonrisitaslabastidaquetehostienmarikon}) into the Euclidean metric (\ref{mekagoentuputakaraluisibanez}), {\it i.e.}, 
\begin{equation}
g_{\rm lin}=f_{\rm hyp}g_{\rm hyp}.
\label{ramayomekagoentuputakara}
\end{equation}
The above conformal transformation to $g_{\rm lin}=\sum_{k=1}^n{\rm d}\bar w^k{\rm d}w^k$ is induced by the change of variables (\ref{queputocalvoeresibanez}) that solves the differential equations
\begin{equation}
{\rm d}w^k=\frac{{\rm d}z^k}{1-\vert\vert z\vert\vert^2}, \qquad {\rm d}\bar w^k=\frac{{\rm d}\bar z^k}{1-\vert\vert z\vert\vert^2}.
\label{kabronpajares}
\end{equation}
The solution to (\ref{kabronpajares}) provides us with a positive function $h_{\rm hyp}(x)$ such that 
\begin{equation}
\vert\vert z\vert\vert^2=h_{\rm hyp}(\vert\vert w\vert\vert^2).
\label{putodelaguila}
\end{equation}

Now let hyperbolic creation and annihilation operators $(A^k_{\rm hyp})^+$ and $A^k_{\rm hyp}$ correspond to the coordinates $\bar z^k$ and $z^k$, respectively. Linear creation and annihilation operators $(A^k_{\rm lin})^+$ and $A^k_{\rm lin}$ respectively correspond to the coordinates $\bar w^k$ and $w^k$ obtained as a solution to (\ref{kabronpajares}). The classical Hamiltonian (\ref{cagonlabastida})
\begin{equation}
H_{\rm hyp}=-\ln\left(1-\vert\vert z\vert\vert^2\right)
\label{luisitoibanezmekagoentuputacara}
\end{equation}
can be reexpressed as
\begin{equation}
H_{\rm hyp}=-\ln\left(1-h_{\rm hyp}(\vert\vert w\vert\vert^2)\right).
\label{barbobquetehostien}
\end{equation}
Quantum--mechanically we make the replacement
\begin{equation}
\vert\vert w\vert\vert^2\mapsto\sum_{k=1}^n\left((A^k_{\rm lin})^+A^k_{\rm lin}+\frac{1}{2}\right),
\label{kirosmarikon}
\end{equation}
so the quantum Hamiltonian operator is
\begin{equation}
H_{\rm hyp}=-\ln\left\{1-h_{\rm hyp}\left(\sum_{k=1}^n\left((A^k_{\rm lin})^{+} A^k_{\rm lin}+\frac{1}{2}\right)\right)\right\}.
\label{hacheproj}
\end{equation}
Diagonalising first the argument of the logarithm as in eqns. (\ref{cesargomezhijoputa}), (\ref{velascocabronazo}), (\ref{velascocabron}), the eigenvalue equation for the hyperbolic Hamiltonian (\ref{hacheproj}) reads 
\begin{equation}
H_{\rm hyp}|m_1,\ldots, m_n\rangle_{\rm hyp} = - \ln\left\{1-h_{\rm hyp}\left(\sum_{k=1}^n 
\left(m_k + {1\over 2}\right)\right)\right\}|m_1,\ldots, m_n\rangle_{\rm hyp},
\label{oeggppv}
\end{equation}
where
\begin{equation}
|m_1,\ldots, m_n\rangle_{\rm hyp} = \frac{1}{\sqrt{m_1!\cdots m_n!}}
\left((A^1_{\rm lin})^+\right)^{m_1}\cdots \left((A^n_{\rm lin})^+\right)^{m_n}\vert 0\rangle_{\rm hyp}.
\label{velaskocabron}
\end{equation}
The occupation numbers $m_k$, $k=1,\ldots, n$, run over all the nonnegative integers, and the Hilbert space ${\cal H}_{\rm hyp}$ is (the closure of) the linear span of all the states $|m_1,\ldots, m_n\rangle_{\rm hyp}$. Thus ${\cal H}_{\rm hyp}$ is infinite--dimensional, in agreement with eqns. (\ref{labastidamariconazo}), (\ref{ramallomecagoentuputamadre}). One could also express quantum states on $B^n$as the result of the action of hyperbolic creation operators $(A^k_{\rm hyp})^+$ on the hyperbolic vacuum $\vert 0\rangle_{\rm hyp}$, at the cost of losing the nice interpretation of eqn. (\ref{velaskocabron}), namely, that each integer power of a creation operator contributes to the state $|m_1,\ldots, m_n\rangle_{\rm hyp}$ by one quantum. 

When $x\to 0$ we have $h_{\rm hyp}(x)\simeq x$. This ensures that, in the limit of small quantum numbers, eqns. (\ref{hacheproj}) and (\ref{oeggppv}) correctly reduce to their expected limits (\ref{cesargomezhijoputa}) and (\ref{velascocabronazo}). This makes sense as, in a neighbourhood of the origin in $B^n$, the hyperbolic oscillator reduces to the linear oscillator, and curvature effects can be neglected. 
The limit of small quantum  numbers is the strong quantum regime. On the contrary, in the limit of large quantum numbers, or classical limit, we have $\vert\vert z\vert\vert\to 1$, $\vert\vert w\vert\vert\to \infty$, so it must hold that $h_{\rm hyp}(x)\to 1$ as $x\to \infty$. Hence the effects of the negative curvature of $B^n$ can no longer be neglected as we approach the boundary of hyperbolic space. The effects of classical curvature due to the logarithm in the K\"ahler potential (\ref{cagonlabastida}) are suppressed, or smoothed out, quantum--mechanically.

\subsection{Projective space}\label{labastidakabron}

Let $Z^1,\ldots, Z^{n+1}$ denote homogeneous coordinates on $\mathbb{CP}^n$. The chart defined by $Z^j\neq 0$ covers one copy of the open set ${\cal U}_j=\mathbb{C}^n$. On the latter we have the holomorphic coordinates $z^k_{(j)}=Z^k/Z^j$, $k=1, \ldots, n+1$, $k\neq j$; there are $n+1$ such coordinate charts. $\mathbb{CP}^n$ is a K\"ahler manifold with respect to the Fubini--Study metric, with constant positive scalar curvature. On 
$({\cal U}_j, z_{(j)}^k)$ the K\"ahler potential reads, dropping the subindex $j$ for simplicity,
\begin{equation}
K_{\rm proj}=\ln{\left(1 + \vert\vert z\vert\vert^2\right)}.
\label{fubst}
\end{equation}
On the same chart, the projective symplectic form is
\begin{equation}
\omega_{\rm proj} = {-{\rm i}\over (1+\vert\vert z\vert\vert^2)^2}\sum_{k=1}^n{\rm d} \bar z^k \wedge {\rm d} {  z}^k,
\label{sonrisitaslabastidacasposo}
\end{equation}
while the Fubini--Study metric reads
\begin{equation}
g_{\rm proj} = {1\over (1+\vert\vert z\vert\vert^2)^2}\sum_{k=1}^n{\rm d} \bar z^k {\rm d} {  z}^k.
\label{sonrisitaslabastidacasposodemierda}
\end{equation}
The Picard group is the additive group of integers \cite{GFH},
\begin{equation}
{\rm Pic}\,(\mathbb{CP}^n)=\mathbb{Z}.
\label{athh}
\end{equation}
The class $l=0$ corresponds to the trivial complex line bundle; all other classes $l\neq 0$ correspond to nontrivial line bundles. 
On $({\cal U}_j, z_{(j)}^k)$, we denote the (fibrewise) generator of the line bundle corresponding to the class $l$ by $\vert 0(j)\rangle_{\rm proj}^l$. For simplicity we will concentrate on the class $l=1$; see ref. \cite{PQM} for the general case.
Then the symplectic volume of $\mathbb{CP}^{n}$ can be normalised as 
\begin{equation}
\int_{\mathbb{CP}^n}\omega_{\rm proj}^n=n+1.
\label{labastidamecagoentuputamadre}
\end{equation}

As explained, we take the classical Hamiltonian function on the coordinate chart $({\cal U}_j, z_{(j)}^k)$ equal to the K\"ahler potential (\ref{fubst}). Let $\pi^{rs}_{\rm proj}$ denote the Poisson tensor corresponding to the symplectic form (\ref{sonrisitaslabastidacasposo}). Now the space of all solutions to Hamilton's equations
\begin{equation}
\dot z^k=\left\{z^k,K_{\rm proj}\right\}= \pi^{rs}_{\rm proj}\frac{\partial z^k}{\partial z^r}\frac{\partial K_{\rm proj}}{\partial \bar z^s}=
\pi^{ks}_{\rm proj}\frac{\partial K_{\rm proj}}{\partial \bar z^s}=-{\rm i}\,z^k\left(1+\vert\vert z\vert\vert^2\right)
\label{putovelascoquetepartaunrayo}
\end{equation}
is in fact the whole coordinate chart $({\cal U}_j, z_{(j)}^k)$. The right--hand side of (\ref{putovelascoquetepartaunrayo}) contains the square root of the conformal factor
\begin{equation}
f_{\rm proj}=\left(1+\vert\vert z\vert\vert^2\right)^2
\label{delaguilaquetepartaunrayo}
\end{equation}
that transforms the Fubini--Study metric (\ref{sonrisitaslabastidacasposodemierda}) into the Euclidean metric (\ref{mekagoentuputakaraluisibanez}), {\it i.e.},
\begin{equation}
g_{\rm lin}=f_{\rm proj}g_{\rm proj}.
\label{labastidatepartolacaramarikondemierda}
\end{equation}
The above conformal transformation to $g_{\rm lin}=\sum_{k=1}^n{\rm d}\bar w^k{\rm d}w^k$ is induced by the change of variables that, on the chart $({\cal U}_j, z_{(j)}^k)$, solves the differential equations
\begin{equation}
{\rm d}w^k=\frac{{\rm d}z^k}{1+\vert\vert z\vert\vert^2}, \qquad {\rm d}\bar w^k=\frac{{\rm d}\bar z^k}{1+\vert\vert z\vert\vert^2}.
\label{kabronazopajares}
\end{equation}
Thus, {\it e.g.}, when $n=1$, this change of variables is given by the usual stereographic projection from the plane to the Riemann sphere.
By the same reasoning as in section \ref{javiermaschupamelapolla}, a positive function $h_{\rm proj}(x)$ exists such that 
\begin{equation}
\vert\vert z\vert\vert^2=h_{\rm proj}(\vert\vert w\vert\vert^2).
\label{putodelaguilamarikondemierda}
\end{equation}
Moreover, $h_{\rm proj}(x)\simeq x$ when $x\to 0$, because the projective oscillator approaches the linear oscillator in this limit. 
This corresponds to dropping the logarithm in the K\"ahler potential (\ref{fubst}).

On the coordinate chart under consideration, $\bar z^k$ and $z^k$ respectively give rise to projective creation and annihilation operators $(A^k_{\rm proj})^+$ and  $A^k_{\rm proj}$ acting on the vacuum $\vert 0\rangle_{\rm proj}^{(l=1)}$. Linear creation and annihilation operators $(A^k_{\rm lin})^+$ and $A^k_{\rm lin}$ correspond to the coordinates $\bar w^k$ and $w^k$, respectively. The classical Hamiltonian (\ref{fubst})
\begin{equation}
H_{\rm proj}=\ln\left(1+\vert\vert z\vert\vert^2\right)
\label{luisitoibanezmekagoentuputasombra}
\end{equation}
can be reexpressed as
\begin{equation}
H_{\rm proj}=\ln\left(1+h_{\rm proj}(\vert\vert w\vert\vert^2)\right).
\label{barbobquetehostienhijoperra}
\end{equation}
Now quantum--mechanically we apply the replacement
\begin{equation}
\vert\vert w\vert\vert^2\mapsto\sum_{k=1}^n\left((A^k_{\rm lin})^+A^k_{\rm lin}+\frac{1}{2}\right),
\label{marianokirosmarikon}
\end{equation}
so the quantum Hamiltonian operator is, on the given chart,
\begin{equation}
H_{\rm proj}=\ln\left\{1+h_{\rm proj}\left(\sum_{k=1}^n\left((A^k_{\rm lin})^{+} A^k_{\rm lin}+\frac{1}{2}\right)\right)\right\}.
\label{velascotontodelculo}
\end{equation}
Proceeding as in previous sections we find
\begin{equation}
H_{\rm proj}|m_1,\ldots, m_n\rangle_{\rm proj}^{(l=1)} = \ln\left\{1+h_{\rm proj}\left(\sum_{k=1}^n 
\left(m_k + {1\over 2}\right)\right)\right\}|m_1,\ldots, m_n\rangle_{\rm proj}^{(l=1)},
\label{kabronvelasko}
\end{equation}
where
\begin{equation}
|m_1,\ldots, m_n\rangle_{\rm proj}^{(l=1)} = \frac{1}{\sqrt{m_1!\cdots m_n!}}
\left((A^1_{\rm lin})^+\right)^{m_1}\cdots \left((A^n_{\rm lin})^+\right)^{m_n}\vert 0\rangle_{\rm proj}^{(l=1)}.
\label{velaskokabron}
\end{equation}
In agreement with eqns. (\ref{labastidamariconazo}), (\ref{labastidamecagoentuputamadre}) there are $n+1$ states, as
the Hilbert space ${\cal H}_{\rm proj}^{(l=1)}$ over the given chart is the linear span of the vectors $\vert m_1, \ldots, m_n\rangle_{\rm proj}^{(l=1)}$, where the occupation numbers are either all zero [for the vacuum $\vert 0\rangle_{\rm proj}^{(l=1)}$], or all are zero but one [for the $p$--th excited state $(A^p_{\rm lin})^+\vert 0\rangle_{\rm proj}^{(l=1)}$, $p=1, \ldots, n$]. With the same {\it caveat}\/ of section \ref{javiermaschupamelapolla}, one could also express quantum states on $\mathbb{CP}^n$ as the result of the action of projective creation operators $(A^k_{\rm proj})^+$ on the projective vacuum $\vert 0\rangle^{(l=1)}_{\rm proj}$. Transition functions for this bundle of Hilbert spaces over $\mathbb{CP}^n$ have been given in ref. \cite{PQM}. The corresponding bundles over $\mathbb{C}^n$ and $B^n$ were trivial due to contractibility.

\subsection{Grassmann manifolds}\label{alvarezgaumemekagoentuputamadremarikonazodeplaya}

Let $G_{r,r'}(\mathbb{C})$ be the complex Grassmann manifold of $r$--dimensional hyperplanes in $\mathbb{C}^{r+r'}$. It can be thought of as
the manifold \cite{KN}
\begin{equation}
G_{r,r'}(\mathbb{C})=U(r+r')/U(r)\times U(r'),
\label{casposilloramallo}
\end{equation}
and it reduces to projective space when $r=1$. In order to obtain holomorphic coordinates on $G_{r,r'}(\mathbb{C})$, let $M(r+r',r;\mathbb{C})$ be the space of complex matrices with $r$ columns, $r+r'$ rows and rank $r$; this is the set of $r$ linearly independent vectors within 
$\mathbb{C}^{r+r'}$. Each set of $r$ linearly independent vectors within $\mathbb{C}^{r+r'}$ determines an $r$--dimensional hyperplane. This gives a natural projection 
\begin{equation}
\Pi\colon M(r+r',r;\mathbb{C})\longrightarrow G_{r,r'}(\mathbb{C}). 
\label{kakapaaralabastida}
\end{equation}
Representing the coordinates on $M(r+r',r;\mathbb{C})$ by $r\times(r+r')$ complex matrices $Z$, we define a closed $(1,1)$--form on $M(r+r',r;\mathbb{C})$ by
\begin{equation}
\tilde\omega=-{\rm i}\,\partial\bar\partial\ln\det (Z^+Z).
\label{kabronlabastidakabron}
\end{equation}
Denote by $\omega_{\rm Gr}$ the projection of $\tilde\omega$ onto the base $G_{r,r'}(\mathbb{C})$, {\it i.e.}, $\Pi^*(\omega_{\rm Gr})=\tilde\omega$. Then $\omega_{\rm Gr}$ is the K\"ahler symplectic form of a K\"ahler metric $g_{\rm Gr}$ on $G_{r,r'}(\mathbb{C})$. Write
\begin{equation}
Z=\left(\begin{array}{c}
Z_0\\
Z_1\end{array}\right),
\label{luisitoluisitocabroncabron}
\end{equation}
where $Z_0$ is an $r\times r$ matrix and $Z_1$ is an $r\times r'$ matrix, and consider the open subset ${\cal U}$ of $G_{r,r'}(\mathbb{C})$ defined by $\det Z_0\neq 0$. Setting $z=Z_1Z_0^{-1}$, we may use the matrix $z$ as a local holomorphic coordinate on ${\cal U}$; there are $\left(r+r'\atop r\right)$ such coordinate charts, which turn $G_{r,r'}(\mathbb{C})$ into a K\"ahler manifold of complex dimension $n=rr'$. Since $Z^+Z=Z_0^+({\bf 1}+z^+z)Z_0$, where ${\bf 1}$ is the $r\times r$ identity matrix, we obtain
\begin{equation}
\partial\bar\partial\ln\det(Z^+Z)=\partial\bar\partial\ln\det({\bf 1}+z^+z).
\label{cabronluisitoalvarezgaumecabron}
\end{equation}
Therefore on $G_{r,r'}(\mathbb{C})$ there are holomorphic coordinate charts $({\cal U}_j, z^{ab}_{(j)})$, where $a=1, \ldots, r$ and $b=1, \ldots, r'$ run over the matrix entries of $z_{(j)}$ and $j$ runs over the $\left(r+r'\atop r\right)$ different charts. For simplicity we drop the subindex $j$. Then we have a K\"ahler potential on $({\cal U}, z^{ab})$
\begin{equation}
K_{\rm Gr}=\ln\det({\bf 1}+z^+z),
\label{ibanezeresunputondemierda}
\end{equation}
with a K\"ahler symplectic form
\begin{equation}
\omega_{\rm Gr}=-{\rm i}\,\partial\bar\partial\left(\ln\det({\bf 1}+z^+z)\right){\rm tr}\,({\rm d}z^+\wedge{\rm d}z),
\label{mierdaparaluisibanez}
\end{equation}
and a K\"ahler metric
\begin{equation}
g_{\rm Gr}=\partial\bar\partial\left(\ln\det({\bf 1}+z^+z)\right){\rm tr}\,({\rm d}z^+{\rm d}z).
\label{quetefollenluisibanez}
\end{equation}
This metric is invariant under $U(r+r')$, which acts transitively on $G_{r,r'}(\mathbb{C})$.
Now the symplectic volume of $G_{r,r'}(\mathbb{C})$ is finite because $G_{r,r'}(\mathbb{C})$ is compact. Let us pick a vacuum 
$\vert 0\rangle_{\rm Gr}^{(l)}$ corresponding to the Picard class $l$, and normalise the symplectic volume of $G_{r,r'}(\mathbb{C})$ as
\begin{equation}
\int_{G_{r,r'}(\mathbb{C})}\omega_{\rm Gr}^n = rr'+1=n+1.
\label{quetefollenlabastidademierda}
\end{equation}
Essentially this choice of a vacuum corresponds, in the language of ref. \cite{PQM}, to the {\it fundamental representation}\/ of the unitary groups in eqn. (\ref{casposilloramallo}); higher representations can be treated as in ref. \cite{PQM}.

Next we take the classical Hamiltonian function on the chart $({\cal U}, z^{ab})$ equal to the K\"ahler potential (\ref{ibanezeresunputondemierda}),
\begin{equation}
H_{\rm Gr}=\ln\det({\bf 1}+z^+z).
\label{ibanezeresunputonazodemierda}
\end{equation}
Letting $\pi^{ab,cd}_{\rm Gr}$ denote the Poisson tensor corresponding to the symplectic form (\ref{mierdaparaluisibanez}), the space of all solutions to Hamilton's equations
\begin{equation}
\dot z^{ab}=\left\{z^{ab}, H_{\rm Gr}\right\}=\pi^{cd,ef}_{\rm Gr}\frac{\partial z^{ab}}{\partial z^{cd}}\frac{\partial H_{\rm Gr}}{\partial \bar z^{ef}}=\pi^{ab,ef}_{\rm Gr}\frac{\partial H_{\rm Gr}}{\partial \bar z^{ef}}
\label{mecagoentuputamadreluisibanez}
\end{equation}
is, on general grounds, the whole coordinate chart $({\cal U}, z^{ab})$. Hence our choice of a Hamiltonian makes sense. There is also a conformal factor 
\begin{equation}
f_{\rm Gr}=\left(\partial\bar\partial\ln\det({\bf 1}+z^+z)\right)^{-1}
\label{velaskokasposo}
\end{equation}
that transforms the metric (\ref{quetefollenluisibanez}) into the Euclidean metric (\ref{mekagoentuputakaraluisibanez}), {\it i.e.},
\begin{equation}
g_{\rm lin}=f_{\rm Gr}g_{\rm Gr}
\label{kakaxvelasko}
\end{equation}
The above conformal transformation to $g_{\rm lin}={\rm tr}\,({\rm d}w^+{\rm d}w)$ is induced by a certain change of variables on the chart $({\cal U},z^{ab})$
\begin{equation}
z^{ab}\rightarrow w^{ab}(z^{cd}, \bar z^{cd}), \qquad \bar z^{ab}\rightarrow \bar w^{ab}(z^{cd}, \bar z^{cd})
\label{mierdaxtodos}
\end{equation}
defined as the solution to a set of differential equations whose specific expression is in general quite involved, as the variables are now matrix--valued. However it suffices to know that the metric (\ref{quetefollenluisibanez}) is conformal to the Euclidean metric on $\mathbb{C}^{n}$, $n=rr'$, to which it reduces in a neighbourhood of the origin. Indeed, in the limit $z^{ab}\to 0$, which is the strong quantum limit, the K\"ahler potential (\ref{ibanezeresunputondemierda}) simplifies to ${\rm tr}\,(z^+z)$, which is the Hamiltonian for the harmonic oscillator on $\mathbb{C}^{n}$. Therefore a positive function $h_{\rm Gr}(x)$ of the real variable $x={\rm tr}\,(w^+w)$ exists, such that $h_{\rm Gr}(x)\simeq x$ in the strong quantum limit $x\to 0$, and such that
\begin{equation}
\det({\bf 1}+z^+z)=1+h_{\rm Gr}({\rm tr}\,(w^+w))
\label{marikonestodos}
\end{equation}
for all $z^{ab}$ on the chart $({\cal U},z^{ab})$. On the latter, $\bar z^{ab}$ and $z^{ab}$ respectively give rise to Grassmannian creation and annihilation operators $(A_{\rm Gr}^{ab})^+$ and $A_{\rm Gr}^{ab}$ acting on the vacuum $\vert 0\rangle_{\rm Gr}^{(l)}$. Linear creation and annihilation operators $(A_{\rm lin}^{ab})^+$ and $A_{\rm lin}^{ab}$ respectively correspond to the coordinates $\bar w^{ab}$ and $w^{ab}$ of eqn. (\ref{mierdaxtodos}). The classical Hamiltonian (\ref{ibanezeresunputonazodemierda}) can be reexpressed on $({\cal U}, z^{ab})$ as
\begin{equation}
H_{\rm Gr}=\ln\left\{1+h_{\rm Gr}\left({\rm tr}\,(w^+w)\right)\right\}.
\label{luisibanezeresunputonazodemierda}
\end{equation}
Now our quantisation prescription reads
\begin{equation}
{\rm tr}\,(w^+w)\mapsto\sum_{a=1}^r\sum_{b=1}^{r'}\left(\left(A_{\rm lin}^{ab}\right)^+A_{\rm lin}^{ba}+\frac{1}{2}\right).
\label{kakaxcasposos}
\end{equation}
The dagger on the right--hand side does not refer to the matrix indices on $\mathbb{C}^{rr'}=\mathbb{C}^n$, but to Hermitian conjugation on 
the Hilbert space of quantum states, which will turn out to be $\mathbb{C}^{n+1}$. It is convenient to trade the double indices corresponding to the $r\times r'$ matrix $z^{ab}$ for a single index $k$ running from 1 to $n=rr'$. In this way, substitution of eqn. (\ref{kakaxcasposos}) into eqn. (\ref{luisibanezeresunputonazodemierda}) yields a quantum Hamiltonian on $({\cal U}, z^{ab})$
\begin{equation}
H_{\rm Gr}=\ln\left\{1+h_{\rm Gr}\left(\sum_{k=1}^n\left(\left(A_{\rm lin}^{k}\right)^+A_{\rm lin}^{k}+\frac{1}{2}\right)\right)\right\}.
\label{labastidaketelametanporculokabronazodemierda}
\end{equation}
Proceeding as in previous sections we find
\begin{equation}
H_{\rm Gr}|m_1,\ldots, m_n\rangle_{\rm Gr}^{(l)} =
\ln\left\{1+h_{\rm Gr}\left(\sum_{k=1}^n 
\left(m_k + {1\over 2}\right)\right)\right\}|m_1,\ldots, m_n\rangle_{\rm Gr}^{(l)},
\label{kabronvelaskodemierda}
\end{equation}
where 
\begin{equation}
|m_1,\ldots, m_n\rangle_{\rm Gr}^{(l)} = \frac{1}{\sqrt{m_1!\cdots m_n!}}
\left((A^1_{\rm lin})^+\right)^{m_1}\cdots \left((A^n_{\rm lin})^+\right)^{m_n}\vert 0\rangle_{\rm Gr}^{(l)}.
\label{velaskokabrondeloskojones}
\end{equation}
In agreement with eqns. (\ref{labastidamariconazo}), (\ref{quetefollenlabastidademierda}) there are $n+1$ states.
The Hilbert space ${\cal H}_{\rm Gr}^{(l)}$ over the given chart is the linear span of the vectors $\vert m_1, \ldots, m_n\rangle_{\rm Gr}^{(l)}$, where the occupation numbers are either all zero [for the vacuum $\vert 0\rangle_{\rm Gr}^{(l)}$], or all are zero but one [for the $p$--th excited state $(A^p_{\rm lin})^+\vert 0\rangle_{\rm Gr}^{(l)}$, $p=1, \ldots, n$].  With the same {\it caveat}\/ of section \ref{javiermaschupamelapolla}, one could also express quantum states on $G_{r,r'}(\mathbb{C})$ as the result of the action of Grassmannian creation operators $(A^k_{\rm Gr})^+$ on the Grassmannian vacuum $\vert 0\rangle^{(l)}_{\rm Gr}$. Transition functions for this bundle of Hilbert spaces over $G_{r,r'}(\mathbb{C})$ are, according to ref. \cite{PQM}, those of the holomorphic tangent bundle to $G_{r,r'}(\mathbb{C})$, directly summed with those for the line bundle whose fibrewise generator is the vacuum.

\subsection{Bounded domains}\label{delaguilaeresuncasposodemierda}

Let $D_{r,r'}(\mathbb{C})$ be the space of $r\times r'$ complex matrices $z^{ab}$ satisfying \cite{KN}
\begin{equation}
{\bf 1}-z^+z>0,
\label{luisibanezqetehostienkabron}
\end{equation}
where the right--hand side means {\it  positive definite}, and ${\bf 1}$ stands for the $r\times r$ identity  matrix. Then $D_{r,r'}(\mathbb{C})$ is a bounded domain within $\mathbb{C}^{rr'}$, coincident with hyperbolic space when $r=1$. The $z^{ab}$ satisfying (\ref{luisibanezqetehostienkabron}) are holomorphic coordinates on all of $D_{r,r'}(\mathbb{C})$.  The latter is K\"ahler with respect to the $U(r,r')$--invariant potential
\begin{equation}
K_{\rm Bd}=-\ln\det\,\left({\bf 1}-z^+z\right).
\label{putocesargomezdemierda}
\end{equation}
In fact it holds that $D_{r,r'}(\mathbb{C})=U(r,r')/U(r)\times U(r')$. The K\"ahler symplectic form is
\begin{equation}
\omega_{\rm Bd}=-{\rm i}\,\partial\bar\partial\left(\ln\det\,\left({\bf 1}-z^+z\right)\right){\rm tr}({\rm d}\bar z\wedge{\rm d}z),
\label{cabronkiros}
\end{equation}
and the K\"ahler metric reads
\begin{equation}
g_{\rm Bd}=\partial\bar\partial\left(\ln\det\,\left({\bf 1}-z^+z\right)\right){\rm tr}({\rm d}\bar z\,{\rm d}z).
\label{mekagoenlosputosdelaautonoma}
\end{equation}
Let $\pi^{ab,cd}_{\rm Bd}$ denote the Poisson tensor corresponding to the symplectic form (\ref{cabronkiros}).  Picking a classical Hamiltonian equal to the K\"ahler potential, 
\begin{equation}
H_{\rm Bd}=-\ln\det\,\left({\bf 1}-z^+z\right),
\label{etadejadematar}
\end{equation}
we have that the energy is bounded from below, and the space of all solutions to Hamilton's equations
\begin{equation}
\dot z^{ab}=\left\{z^{ab}, H_{\rm Bd}\right\}=\pi^{cd,ef}_{\rm Bd}\frac{\partial z^{ab}}{\partial z^{cd}}\frac{\partial H_{\rm Bd}}{\partial \bar z^{ef}}=\pi^{ab,ef}_{\rm Bd}\frac{\partial H_{\rm Bd}}{\partial \bar z^{ef}}
\label{mecagoentuputamadreluisibanezcasposodemierda}
\end{equation}
is, on general grounds, the whole bounded domain $D_{r,r'}(\mathbb{C})$.

Now the metric (\ref{mekagoenlosputosdelaautonoma}) is conformal to the Euclidean metric on $\mathbb{C}^{rr'}$, 
\begin{equation}
g_{\rm lin}=f_{\rm Bd}g_{\rm Bd},
\label{mekagoenelputocesargomezdemierda}
\end{equation}
with a conformal factor 
\begin{equation}
f_{\rm Bd}=\left(\partial\bar\partial\ln\det\,\left({\bf 1}-z^+z\right)\right)^{-1}.
\label{kurtalechneramoremio}
\end{equation}
The above conformal transformation to $g_{\rm lin}={\rm tr}\,({\rm d}w^+{\rm d}w)$ is induced by a certain change of variables 
\begin{equation}
z^{ab}\rightarrow w^{ab}(z^{cd}, \bar z^{cd}), \qquad \bar z^{ab}\rightarrow \bar w^{ab}(z^{cd}, \bar z^{cd})
\label{mierdaxtodalaeta}
\end{equation}
defined as the solution to a set of differential equations whose specific expression is in general quite involved, as the variables are again matrix--valued. However it suffices to know that the metric (\ref{mekagoenlosputosdelaautonoma}) reduces to the Euclidean metric on 
$\mathbb{C}^{n}$, $n=rr'$, in a neighbourhood of the origin. Indeed, in the limit $z^{ab}\to 0$, which is the strong quantum limit, the K\"ahler potential (\ref{putocesargomezdemierda}) simplifies to ${\rm tr}\,(z^+z)$, which is the Hamiltonian for the harmonic oscillator on $\mathbb{C}^{n}$. Therefore a positive function $h_{\rm Bd}(x)$ of the real variable $x={\rm tr}\,(w^+w)$ exists, such that $h_{\rm Bd}(x)\simeq x$ in the strong quantum limit $x\to 0$, and such that
\begin{equation}
\det({\bf 1}-z^+z)=1-h_{\rm Bd}({\rm tr}\,(w^+w)),
\label{marikonestodoslosvascos}
\end{equation}
with the right--hand side always positive. From the variables $\bar z^{ab}$ and $z^{ab}$ we can construct bounded--domain creation and annihilation operators, $(A_{\rm Bd}^{ab})^+$ and $A_{\rm Bd}^{ab}$ respectively. Linear creation and annihilation operators $(A_{\rm lin}^{ab})^+$ and $A_{\rm lin}^{ab}$ respectively correspond to the coordinates $\bar w^{ab}$ and $w^{ab}$ of eqn. (\ref{mierdaxtodalaeta}), in terms of which the classical Hamiltonian (\ref{etadejadematar}) can be reexpressed as
\begin{equation}
H_{\rm Bd}=-\ln\left(1-h_{\rm Bd}\left({\rm tr}\,(w^+w)\right)\right).
\label{vascoscabrones}
\end{equation}
Now our quantisation prescription reads
\begin{equation}
{\rm tr}\,(w^+w)\mapsto\sum_{a=1}^r\sum_{b=1}^{r'}\left(\left(A_{\rm lin}^{ab}\right)^+A_{\rm lin}^{ba}+\frac{1}{2}\right).
\label{kakaxcaspososetarras}
\end{equation}
The dagger on the right--hand side does not refer to the matrix indices on $\mathbb{C}^{rr'}=\mathbb{C}^n$, but to Hermitian conjugation on 
the Hilbert space of quantum states, which will turn out to be infinite--dimensional. This is a consequence of the infinite symplectic volume of $D_{r,r'}(\mathbb{C})$,
\begin{equation}
\int_{D_{r,r'}(\mathbb{C})}\omega_{\rm Bd}^n=\infty.
\label{bastaya}
\end{equation}
Moreover $D_{r,r'}(\mathbb{C})$ is contractible as a manifold, so its Picard group is trivial. Hence all holomorphic line bundles over $D_{r,r'}(\mathbb{C})$ are trivial; we denote their fibrewise generator by $\vert 0\rangle_{\rm Bd}$. 

It is convenient to trade the double indices corresponding to the $r\times r'$ matrix $z^{ab}$ for a single index $k$ running from 1 to $n=rr'$. In this way, substitution of eqn. (\ref{kakaxcaspososetarras}) into eqn. (\ref{vascoscabrones}) yields a quantum Hamiltonian 
\begin{equation}
H_{\rm Bd}=-\ln\left\{1-h_{\rm Bd}\left(\sum_{k=1}^n\left(\left(A_{\rm lin}^{k}\right)^+A_{\rm lin}^{k}+\frac{1}{2}\right)\right)\right\}.
\label{labastidaketelametanporculokabronazodemierdaxti}
\end{equation}
Proceeding as by now usual we find
\begin{equation}
H_{\rm Bd}|m_1,\ldots, m_n\rangle_{\rm Bd} =
-\ln\left\{1-h_{\rm Bd}\left(\sum_{k=1}^n 
\left(m_k + {1\over 2}\right)\right)\right\}|m_1,\ldots, m_n\rangle_{\rm Bd},
\label{kabronotegidemierda}
\end{equation}
where 
\begin{equation}
|m_1,\ldots, m_n\rangle_{\rm Bd} = \frac{1}{\sqrt{m_1!\cdots m_n!}}
\left((A^1_{\rm lin})^+\right)^{m_1}\cdots \left((A^n_{\rm lin})^+\right)^{m_n}\vert 0\rangle_{\rm Bd}.
\label{otegikabrondeloskojones}
\end{equation}
The occupation numbers $m_k$, $k=1,\ldots, n$, run over all the nonnegative integers, and the Hilbert space ${\cal H}_{\rm Bd}$ is (the closure of) the linear span of all the states $|m_1,\ldots, m_n\rangle_{\rm Bd}$. Thus ${\cal H}_{\rm Bd}$ is infinite--dimensional, in agreement with eqns. (\ref{labastidamariconazo}), (\ref{bastaya}). With the same {\it caveat}\/ of section \ref{javiermaschupamelapolla}, one could also express quantum states on $D_{r,r'}(\mathbb{C})$ as the result of the action of bounded--domain creation operators $(A^k_{\rm Bd})^+$ on the bounded--domain vacuum $\vert 0\rangle_{\rm Bd}$. 
\vskip1cm

{\bf Acknowledgements}

It is a great pleasure to thank J. de Azc\'arraga for encouragement and support. Technical discussions with U. Bruzzo and M. Schlichenmaier are gratefully acknowledged. The author thanks S. Theisen and Max-Planck-Institut f\"ur Gravitationsphysik, Albert-Einstein-Institut,
where this work was completed, for hospitality. This work has been partially supported by research grant BFM2002--03681 from Ministerio de Ciencia y Tecnolog\'{\i}a, by EU FEDER funds, by Fundaci\'on Marina Bueno and by Deutsche Forschungsgemeinschaft.

\end{document}